\documentclass[manuscript]{acmart}
\usepackage{array}
\usepackage{ragged2e}
\usepackage{colortbl}
\usepackage{color}
\usepackage{tabularray}
\usepackage{caption}
\usepackage{subcaption}
\usepackage{geometry}
\definecolor{Mercury}{rgb}{0.901,0.898,0.898}

\AtBeginDocument{%
  }
\setcopyright{acmlicensed}
\copyrightyear{2025}
\acmYear{2025}
\acmDOI{XXXXXXX.XXXXXXX}
\acmConference[CHI '25]{...}{...}{...}
\acmISBN{...}

\begin{document}

\title{Practicing Stress Relief for the Everyday: Designing Social Simulation Using VR, AR, and LLMs}

\author{Anna Fang}
\affiliation{%
  \institution{Carnegie Mellon University}
   \country{USA}
}

\author{Hriday Chhabria}
\authornote{Both authors contributed equally to this work.}  

\author{Alekhya Maram}
\authornotemark[1]
\affiliation{%
  \institution{University of Michigan}
  \country{USA}
}
\affiliation{%
  \institution{Barnard College}
  \country{USA}
}

\author{Haiyi Zhu}
\affiliation{%
  \institution{Carnegie Mellon University}
   \country{USA}
}
\renewcommand{\shortauthors}{Fang et al.}

\begin{abstract} 
Stress is an inevitable part of day-to-day life yet many find themselves unable to manage it themselves, particularly when professional or peer support are not always readily available. As self-care becomes increasingly vital for mental well-being, this paper explores the potential of social simulation as a safe, virtual environment for practicing stress relief for everyday situations. Leveraging the immersive capabilities of VR, AR, and LLMs, we developed eight interactive prototypes for various everyday stressful scenarios (e.g. public speaking) then conducted prototype-driven semi-structured interviews with 19 participants. We reveal that people currently lack effective means to support themselves through everyday stress and found that social simulation fills a gap for simulating real environments for training mental health practices. We outline key considerations for future development of simulation for self-care, including risks of trauma from hyper-realism, distrust of LLM-recommended timing for mental health recommendations, and the value of accessibility for self-care interventions.
\end{abstract}


\begin{teaserfigure}
  \includegraphics[width=\textwidth]{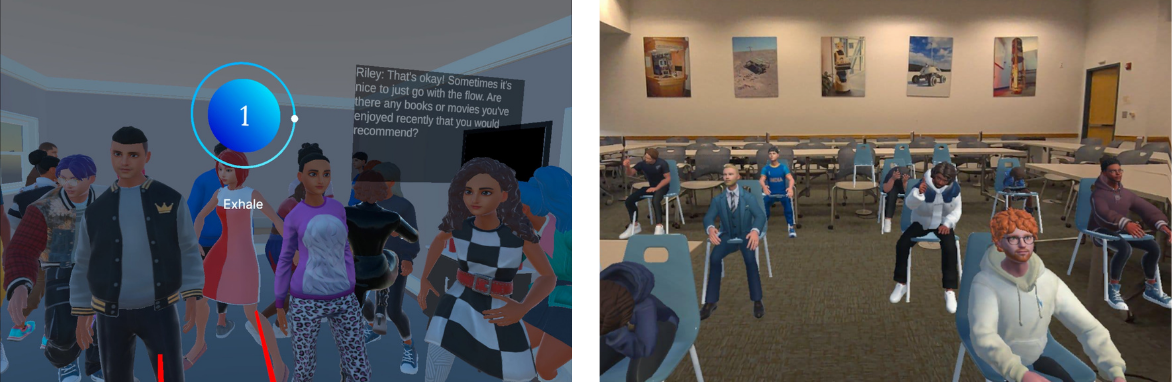}
  \caption{We show participants XR and LLM-powered prototypes that simulate everyday stressors to understand perspectives on simulation for self-care. Pictured are a virtual reality simulation with breathwork guidance for practicing a social party (left) as well as an augmented reality scene for practicing public speaking (right).}
  \label{fig:teaser}
\end{teaserfigure}
\maketitle

\section{Introduction}

When faced with stress and anxiety, taking care of ourselves is challenging. At least 70\% of Americans experience physical and mental symptoms of stress and 58\% of people aged 18 to 34 feel most days contain an overwhelming amount of stress. However, only 37\% of people believe they manage this stress effectively \cite{stressinAmerica}. Although mental health resources have become more accessible through advancements like virtual therapy and online communities \cite{chester2006online, kauer2014online}, most people do not have constant access to these resources when going about their personal, work, or social lives and inevitably encounter stress or even panic. Stress is not only difficult to feel, but also affects our ability to fulfill other aspects of our lives as family members, friends, professionals, and community members \cite{maslow1943theory,world2022guideline}.

Human-Computer Interaction (HCI) research has extensively explored mental health care needs and systems over the past few decades \cite{balcombe2022human}. Despite the many people who are not able to manage their own stress effectively, though, there continues to be a notable gap in applying HCI technology to practicing everyday self-care. One potential avenue for offering people a safe environment to practice their stress relief skills is with \textit{simulation technology} -- technological systems that aim to replicate real-world behaviors, agents, or processes in a virtual environment \cite{hollan1984steamer, tambe1995intelligent, park2022social}. Virtual simulation can aid in stress-relief practices by providing immersive, controlled scenarios that allow individuals to safely engage with and manage stressors. By recreating realistic social and environmental dynamics, users can potentially practice coping mechanisms in a low-risk environment, a preventative self-care measure that can improve emotional regulation and preparedness for real-life situations \cite{carson1998stress}.

In our work, we leverage virtual reality (VR), augmented reality (AR), and large language models (LLMs) to develop social simulation in order to simulate common everyday environments for users to practice self-soothing -- specifically, box breathing, which is a technique well-documented in literature for its effectiveness in treating stress, anxiety, and panic \cite{CLARK198523}. Using eight low fidelity prototypes, we conducted \textit{prototype-driven interviews} with 19 participants to explore their reactions to using simulation for practicing stress relief for the everyday. Specifically, our work explores the research question: 

\indent \textbf{RQ: What are the needs, risks, and design considerations in using simulation technology to help practice management of everyday stress?} 

Our findings indicate that many people lack effective methods for self-care and found our simulation prototypes useful for building skills to deal with everyday stressors; most expressed they would incorporate this use of simulation into their real-world practices as a self-care practice tool. We also studied different dimensions for designing social simulation for self-care, including modality, interactivity, and mental health guidance features. Participants particularly preferred AR due to its integration with real surroundings, which made it easier to transfer skills to real-world contexts and engage physically with the environment. However, participants also saw accessibility as a valuable asset of LLM-powered, purely text-based simulation despite lower immersion, given its likely greater accessibility on-the-go. Our interviews also uncovered trade-offs in designing social simulation for self-care, such as strong preferences for more realism alongside concerns about potential trauma from hyper-realistic designs. Finally, our work highlights key design considerations for using LLMs in social simulation, such as users' concerns about the effects of LLM-generated recommendations on their sense of agency and real-world transferability of skills.
\section{Related Work}
In this section, we review relevant literature to our work, including HCI for mental health, simulation technology, extended reality for mental health, and large language models for mental health.

\subsection{HCI for Mental Health and Self-Care}
HCI work for mental and emotional health has spanned decades of research, ranging from computational approaches involving machine learning and natural language processing \cite{thieme2020machine, balcombe2022human}, to more qualitative understandings of people's needs in digital mental health \cite{pretorius2020searching,lattie2020designing,thieme2016challenges}, to development of technical HCI systems involving wearables and VR interventions \cite{freeman2017virtual,deighan2023social,hickey2021smart}. There is immense potential (and proven success) that technology for mental health care can have. However, most work in HCI scholarship has focused on how to deliver support from a professional therapeutic standpoint, such as internet-based cognitive-behavioral therapy and dialectic therapy \cite{karyotaki2017efficacy, mohr2017intellicare, schroeder2018pocket}, or online peer and social support \cite{yang2024makes,sharma2018mental, andalibi2016understanding}. Our work fills a different need in support, focusing on helping individuals practice skills to \textit{self}-care, particularly during their day-to-day encounters.

Apart from extended reality and mindfulness, which we review below, research in HCI that has studied how people self-care or the efficacy of self-care tools has primarily centered on tracking and self-management for chronic, diagnosable conditions. For example, work has studied how people navigate and manage their own chronic depression through reliance on social ties \cite{burgess2019think} and built systems for people to track their lifestyle habits for bipolar disorder \cite{bardram2013designing}. Additionally, self-care research has primarily centered on \textit{physical health} conditions. For example, a review paper in 2015 that studied self-care technologies in HCI \cite{nunes2015self} regarding chronic conditions reviewed 29 papers in HCI, where just three included any mental health conditions (bipolar disorder and unipolar depression), while a review done in 2020 on self-care measures for adolescents reviewed health needs for diabetes, cystic fibrosis, and female care without any mental health papers included. Regardless, some aspects of this past literature may still be applicable to self-care for the everyday stress and anxiety people experience, such as the need for accessibility of self-management tools \cite{nunes2015self}.

\subsection{Simulation in HCI}
Agent-based simulation has long been used in the social sciences to explain and test different aspects of human behavior \cite{tesfatsion2006handbook, gilbert2005simulation}. Simulation in HCI have traditionally focused on virtual behaviors in online networks \cite{alvarez2016network, plikynas2015agent, schweitzer2010agent}, or used agent-based simulation to make design decisions for online communities \cite{ren2010agent, ren2014agent}. In recent years, accurate human dialogue and complex social dynamics has also been able to be replicated using simulation techniques through the rise of large language models and generative AI \cite{park2022social, shaikh2024rehearsal}; for example, Park et al.'s work was able to use LLM-powered "generative agents" who act as believable human-like avatars in a virtual town engaging in everyday behaviors like forming interpersonal relationships and coordinating plans \cite{park2023generative}.  

One significant advantage of agent-based simulations, compared to direct experimentation, is their ability to identify downstream and unintended effects in a safe, controlled environment \cite{ren2010agent, ren2014agent}. For example, it can be used to test new algorithms safely before deployment to a real-world community, which is particularly important for the potentially vulnerable and sensitive context of health \cite{liu2023agent}. Since simulations can accurately mirror real-world processes without the associated risks, they provide a safer space for individuals to practice self-care skills at their own pace and in a more comfortable atmosphere. 

\subsection{Immersive Technology for Mental Health}
\subsubsection{Virtual and Augmented Reality Technology for Mental Health and Self-Care}
Among the wide range of technologies that have been leveraged to aid people's mental health, such as chatbots \cite{bae2021social} and wearables \cite{crivelli2019supporting}, extended reality (XR) including both VR and AR has emerged as perhaps the most rapidly growing and promising tool. VR interventions in particular have shown effectiveness in helping people develop sensory awareness and flow \cite{seabrook2020understanding, bruggeman2018hiatus}, conduct mindfulness and meditation practices \cite{kaplan2021impact, waller2021meditating, yildirim2020efficacy}, and to treat a range of mental health conditions \cite{balcombe2022human,botella2017recent,reger2016randomized,falconer2016embodying,oprics2012virtual, meyerbroker2010virtual,freeman2008studying}. 

Apart from treatment for particular conditions, HCI and XR research has also studied mental health interventions for the everyday, novice user. This is perhaps most apparent in HCI's focus on meditation and mindfulness through XR \cite{wang2022reducing}; researchers have used VR and AR to deliver learning curriculuum to teach meditation skills \cite{feinberg2022zenvr}, practice varying breathing techniques \cite{prpa2018attending, wang2023breathero}, help pedestrians navigate more mindfully \cite{chung2016mindful}, and designed multimodal experiences to facilitate walking meditation \cite{tan2023mindful}. XR also allows valuable embodiment of emotions and thoughts into physical space, allowing users to "physically" interact such as punching away negative thoughts \cite{grieger2021trash}. Given that our work focuses on environmental and social simulation of scenarios like public speaking or social anxiety, we also drew from past work studying the effectiveness of XR interventions for accurately simulating anxiety (even for non-anxious people) \cite{maue2022hopohopo} and its ability to effectively reduce stress for experiences like public speaking \cite{lim2023meta, harris2002brief}. However, we note a distinction in our work from past XR systems for mindfulness that have removed people from real-world environments altogether, directing them to calming settings like abstract environments \cite{ng2023virtual}, natural environments \cite{chandrasiri2020virtual}, or otherworldly spaces \cite{miller2023awedyssey}. Overall, XR for self-care has been a rich field of study in HCI, and shown immense promise in its ability to improve symptoms and help people more tangibly interact with their own mental and emotional states.

\subsubsection{Large Language Models (LLMs) for Mental Health}
There has been remarkable success and growth with applying large language models to the mental health domain whether in the clinical setting or non-professional context. LLMs have been used to directly provide therapeutic help \cite{lai2023supporting, loh2023harnessing, fu2023enhancing}, classify mental health experiences \cite{xu2024mental}, and teach social and coping skills \cite{shaikh2024rehearsal, yang2024social, hu2024grow}. There is also emerging work on the capabilities of LLMs to help people gain mental health insights for themselves. For example, LLMs can help patients with their journaling content and then summarize patients' thoughts for their caretakers \cite{kim2024mindfuldiary} and help suggest restructuring of negative thoughts \cite{sharma2024facilitating}.

LLMs have been used to power conversational agents and mental health chatbots for both care-seekers and care-providers. LLM-powered agents have shown they are able to provide context-aware responses that can be helpful for providing mental health support directly to people, such as through providing direct treatment \cite{kian2024can} and helping reframe thought processes \cite{maddela2023training}. LLM-based chatbots can also offer guidance to support-providers such as helping them provide better responses and in diagnosis \cite{sharma2023human, cheng2023now}. However, there are also dangers of LLMs being used in digital mental health intervention, including delivering of misinformation and exhibiting bias \cite{zhou2023synthetic, de2023benefits}. Given these risks, we focus on moving forward in building simulation for self-care in human-centered ways through first understanding human needs, preferences, and boundaries when encountering these technologies. This study probes users' thoughts on LLM-powered guidance, but does not use LLMs to provide direct mental health guidance advice; instead, we restrict LLMs to producing realistic conversational dialogue \cite{ma2023understanding} and gather user reactions to LLM-recommended timing for delivering breathing guidance instead.

Our work lies at the intersection of the above fields. Specifically, we adopt a simulation-based approach, leveraging XR and LLMs to explore users' opinions on being placed in real-world environments to practice self-coping and self-care skills. Our study differentiates itself from prior literature by \textit{applying simulation to teach self-care practices}, and moreover focuses on people's \textit{everyday routine stressors} rather than those with specific conditions.
\section{Methods}

In this section, we describe the three key design dimensions - \textbf{modality}, \textbf{interactivity}, and \textbf{guidance needs} (3.1) - that guide our prototype development (3.2). Using these prototypes, we conducted \textit{\textbf{prototype-driven interviews}} (3.3) and analyzed these interviews using thematic analysis (3.4). 

Our prototype development uses an approach akin to parallel prototyping \cite{dow2010parallel} in order to generate richer insights into people's design preferences and lead to higher quality design recommendations. Given our exploratory goals to uncover users' needs and constraints, we designed eight different prototypes rather than testing one "optimal" solution so we could thoroughly explore the design space. While we did probe participants after each prototype (e.g. \textit{"What are your initial impressions?"}, \textit{"Would you use this?"}), we also focused on asking overarching thoughts and design feedback after participants had finished seeing all eight prototypes (e.g. "\textit{Which of these designs do you prefer, if any, and why?}", "\textit{How realistic did these simulations feel?}").  

\subsection{Key Design Dimensions}
We designed prototypes following three main dimensions guided by our research question and past studies in immersive technology \cite{ryan1994immersion}:

\textbf{Modality}: We aimed to investigate how the form of the simulation — whether VR, AR, or text-based — and thus its level of immersion \cite{ryan1994immersion} affects users’ perceptions of the technology in managing stress.

\textbf{Interactivity}: We also ask what level and types of interaction users desire within these simulations -- whether through dialogue, body language, or environmental feedback. Our prototypes range from non-interactive, static environments to fully interactive settings where users engage in real-time, responsive conversations with LLM-powered avatars. 

\textbf{Guidance Needs}: Our study explores how to help people practice active soothing mechanisms, such as breathing, during moments of high stress. However, some past research in VR/AR has found success in just facilitating exposure (without any incorporated self-care techniques) to help build resilience \cite{krijn2004virtual}. We explore how users felt practicing stress relief techniques superimposed on the simulation of real-world environments; in particular, we selected box breathing as our guidance intervention given its positive effects for stress and relative simplicity to learn for new users \cite{dar2022virtual,bentley2023breathing, boxbreathing}. As a result, for each modality we include prototypes without any breathing intervention and those with intervention (\textbf{\texttt{VR-Guided}}, \textbf{\texttt{AR-Guided}}, \textbf{\texttt{Text-Guided}}).

Through these three dimensions, we designed our prototypes with the goal of providing insights into how different aspects of simulation can contribute to social simulation for self-care. 

\subsection{Design Prototypes}
Although everyone experiences everyday stress \cite{Kingdon_2009}, what triggers that stress varies from person-to-person. To better understand participants’ feelings about our prototypes in the context of their own personal experiences, we designed prototypes around \textbf{three distinct scenarios} that have been simulated in prior research, with the aim for at least one to resonate with participants as a stressor. Due to interview time constraints given our many prototypes, we did not show participants interactive controls for AR and instead asked participants to imagine given the system they had just previously interacted with in VR. Participants were also shown screenshots of prototypes to better visualize the various scenarios and interfaces.

\begin{enumerate}
\item \textbf{Public Speaking Q\&A} \cite{stupar2017beat, slater1999public}. Public speaking is the most common fear \cite{dwyer2012public}. Given the practical challenge in asking participants to give a full speech, we opted to imagine themselves giving a speech and interact in a post-talk Q\&A setting instead. In this scenario, the user is asked to imagine they have just finished giving a speech on any topic of their choosing and answer questions from a 12-person audience. 
\item \textbf{Social Party} \cite{owens2015can,horigome2020virtual}. We selected to simulate a crowded social situation given that social anxiety is the most common anxiety disorder \cite{stein2008social}. In this scenario, the user is asked to imagine they just moved to a new area and have been invited to a party. They are met with a crowded room of people and are greeted in conversation with 4 other party guests.
\item \textbf{Interpersonal Conflict} \cite{shaikh2024rehearsal}. We simulated interpersonal conflict (in this case, with a roommate) due to most people having encountered it but lacking the opportunity to practice the complex dynamics of conflict resolution in an emotionally safe setting \cite{deutsch1994constructive, shaikh2024rehearsal}. Stress during conflict has also been shown to be a barrier to resolution \cite{sillars1982stress}. In this scenario, the user is asked to engage in conflict resolution with their roommate about a topic of their choosing (see Figure \ref{fig:workflow}).
\end{enumerate}

\begin{figure}[!t]
  \begin{tabular}[b]{cc}
    \begin{tabular}[b]{c}
      \begin{subfigure}[b]{0.42\columnwidth}
        \includegraphics[width=\textwidth]{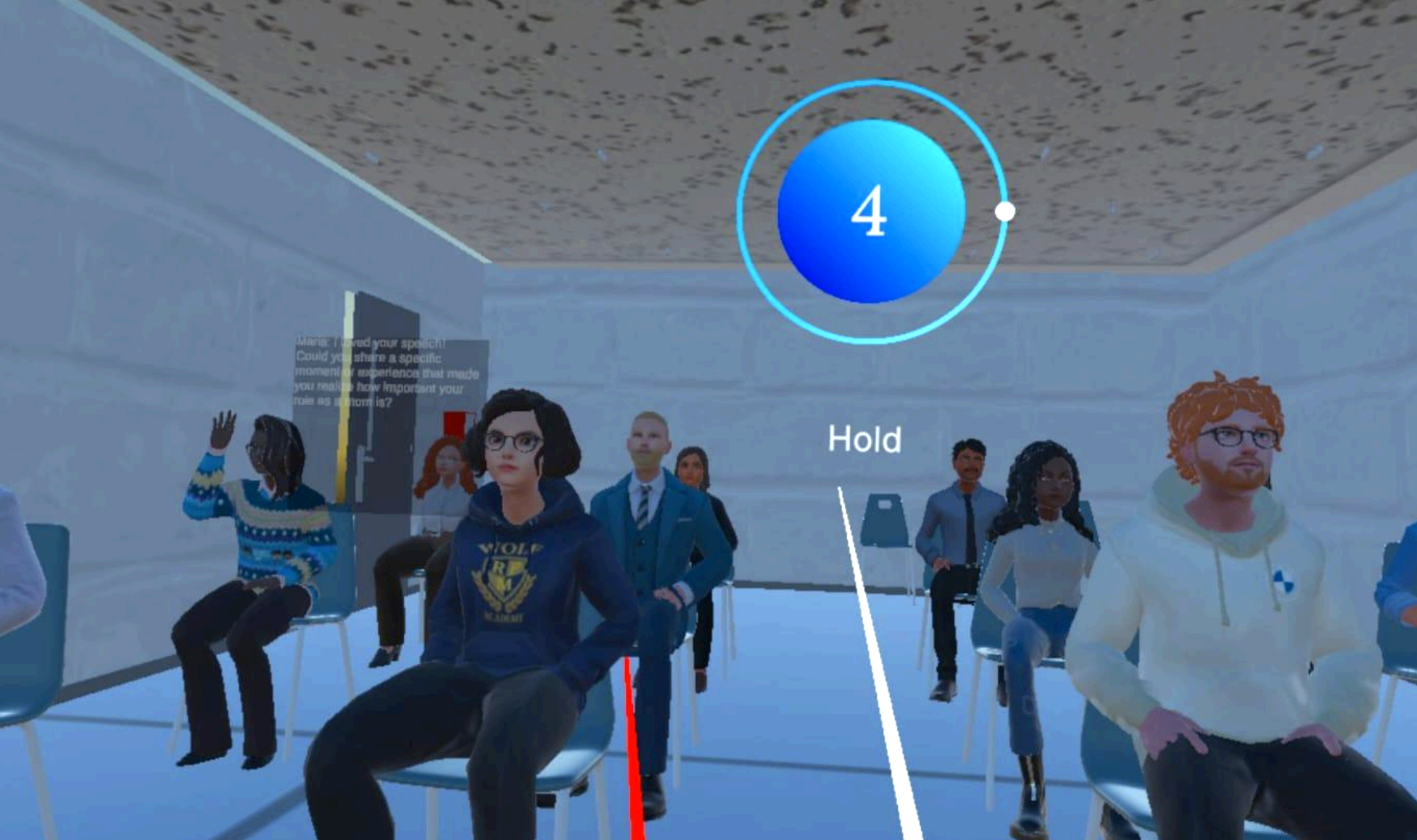}
        \caption{VR-Guided for public speaking scenario.}
      \end{subfigure}\\
      \begin{subfigure}[b]{0.42\columnwidth}
        \includegraphics[width=\textwidth]{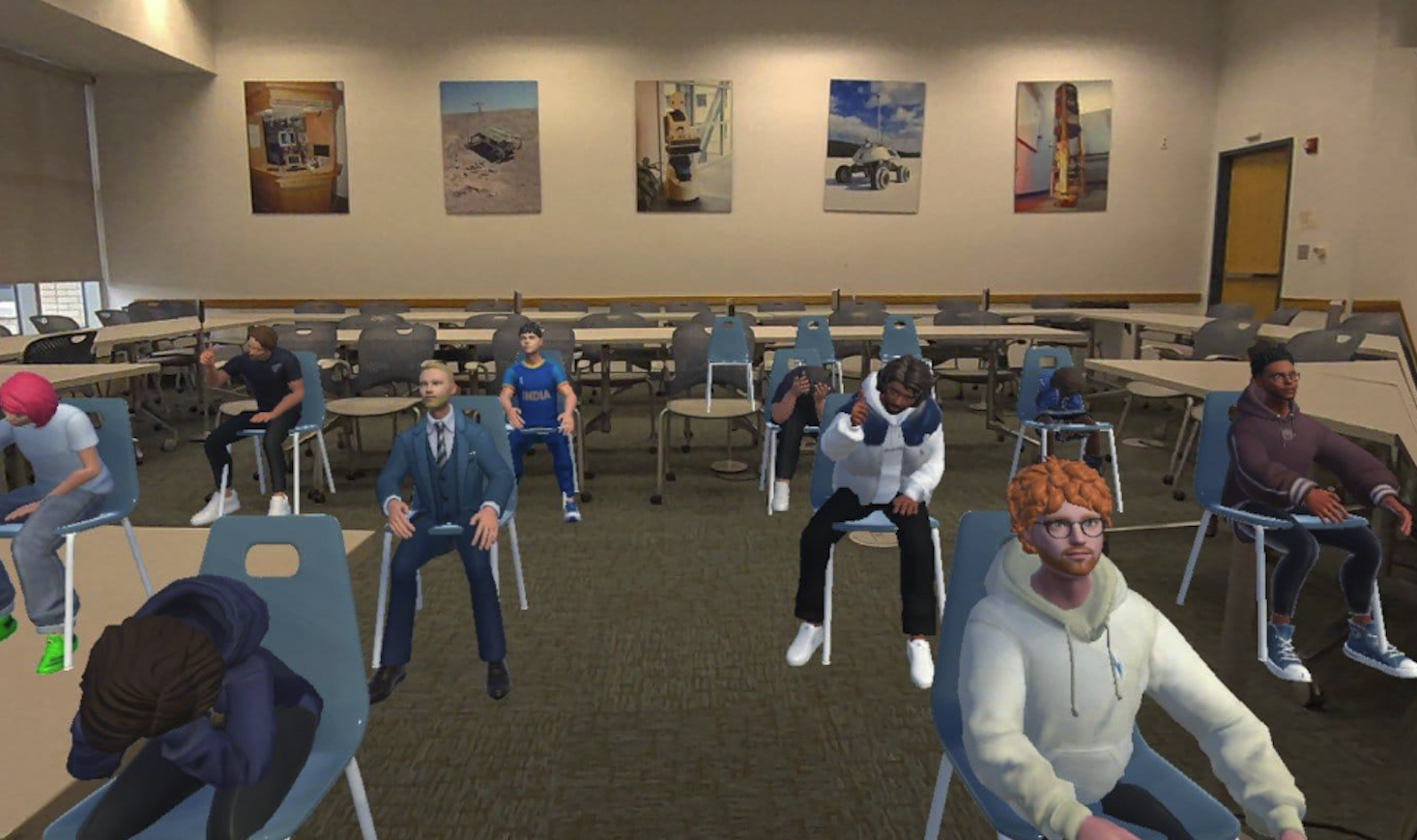}
        \caption{AR-Static for public speaking.}
      \end{subfigure}
    \end{tabular}
    &
    \begin{subfigure}[b]{0.42\columnwidth}
      \includegraphics[width=\textwidth]{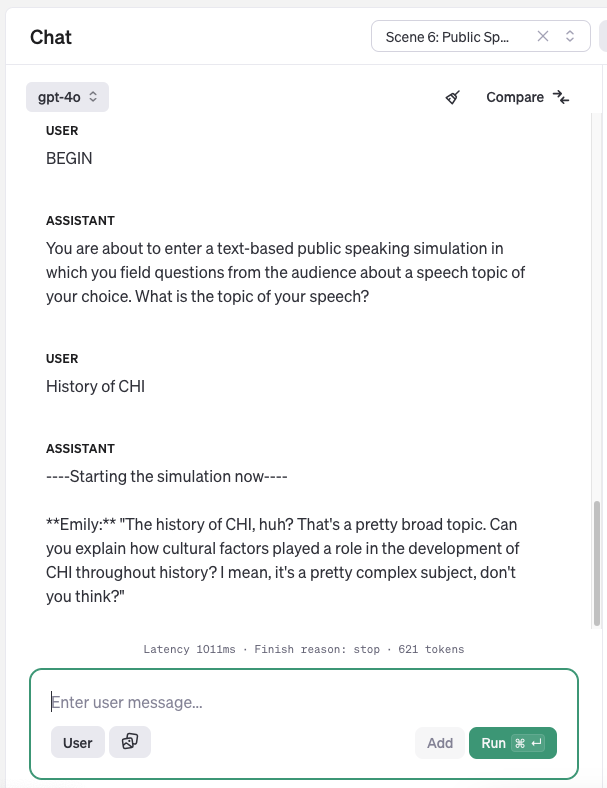}
      \caption{Text-Interactive for public speaking.}
    \end{subfigure}
  \end{tabular}
  \caption{Examples of different prototypes, shown here for the public speaking scenario.}
  \label{fig:ABC}
\end{figure}

Based on our design dimensions of modality, interactivity, and guidance needs, we created the following eight prototypes for each of the three scenarios. We go into more detail on the implementation in the following sections. Example screenshots are shown in Figure \ref{fig:ABC}.

\begin{itemize}  
        \item [1] \textbf{\texttt{VR-Static}}: a non-interactive scene in virtual reality. Avatar(s) and the user do not engage in any dialogue interaction.
        \item [2] \textbf{\texttt{VR-Interactive}}: same as VR-Static, except avatar(s) and the user can engage in dialogue. User can pull up a keyboard to type or dictate their replies.
        \item [3] \textbf{\texttt{VR-Guided}}: same as VR-Interactive, with the addition that the user can press a button on their controller at any point to pull up a breathing guidance orb. The user can use the breathing guidance at any point for as long as they wish.
        \item [4] \textbf{\texttt{AR-Static}}: a non-interactive scene.
        \item [5] \textbf{\texttt{AR-Interactive}}: same as AR-Static, except avatar(s) and the user can engage in dialogue similar to the VR version.
        \item [6] \textbf{\texttt{AR-Guided}}: same as AR-Interactive, with the addition that the user can press a button on their controller at any point to pull up a breathing guidance orb. The user can use the breathing guidance at any point for as long as they wish.
        \item [7]\textbf{\texttt{Text-Interactive}}: users interact with a text-based bot that role plays as an audience/party guests/a roommate, depending on the users' scenario.
        \item [8] \textbf{\texttt{Text-Guided}}: same as Text-Interactive, with the addition of text output of breathing guidance (i.e. "Breathing Guidance: Inhale for 4 seconds. Hold for 4 seconds. Exhale...").

\end{itemize}

\subsubsection{Prototype Development}
Our method involves developing usable, interactive prototypes, as opposed to using storyboards, in order to generate more detailed insights as participants could tangibly experience the immersive capabilities of social simulation. We go into more detail below on building the prototypes, including the visual and audio components of VR/AR and the LLM-powered dialogue.

\textbf{Environment.} For all 18 VR/AR builds (three VR and three AR, across three different scenarios) we built the prototypes using the Unity game engine and deployed on a Meta Quest 3 headset. All avatars were created using Ready Player Me\footnote{https://readyplayer.me/} where the first three authors selected diverse avatars and appropriate clothing for the scenario (e.g. casual outfits for the social party). Additionally, we looped sound clips that suited each scenario using freely available audio from online. Specifically, we included soft classroom and hallway noises for the public speaking scenario, incoherent talking and dance music for the social party scenario, and layered air conditioning and street noises for the interpersonal conflict scenario.

\textbf{Dialogue.} Dialogue for all prototypes is generated using GPT-4o, the most recent GPT version at the time of this study. Our work contributes to the existing but little work on using LLMs to generate realistic avatar dialogue for interactive extended reality \cite{shoa2023sushi, wan2024building}. We used consistent prompting patterns across text-based, VR, and AR prototypes (see Appendix). To ensure natural interactions, the first three authors iterated on prompting patterns to create realistic exchanges. For instance, we asked GPT to generate character personas with seed examples of conversational styles to create realistic and relevant interactions. We also ensured that GPT would output dialogue from individual characters so they spoke one-at-a-time in order to not overwhelm the user. For VR/AR prototypes, we used the OpenAI-Unity package\footnote{https://github.com/srcnalt/OpenAI-Unity} to integrate GPT-4o responses to avatars in VR/AR. For the text-based prototypes, participants interacted directly with GPT-4 via the OpenAI Playground\footnote{https://platform.openai.com/playground}. We conducted pilot testing and gathered feedback from a psychiatry practitioner before proceeding with participant interviews.

\textbf{Interaction and Controls.} The text-based prototypes were created and shown to participants in OpenAI Playground, so participants interacted with GPT directly using a laptop keyboard. For VR/AR interaction, participants interacted with avatars by using the Quest 3 controllers to type on a VR keyboard, or could use speech-to-text dictation that we implemented using WhisperAI\footnote{https://openai.com/index/whisper/}. Participants are stationary in the simulation, although they can turn to look anywhere 360-degrees.

\begin{figure}[t!]
\includegraphics[width=\textwidth]{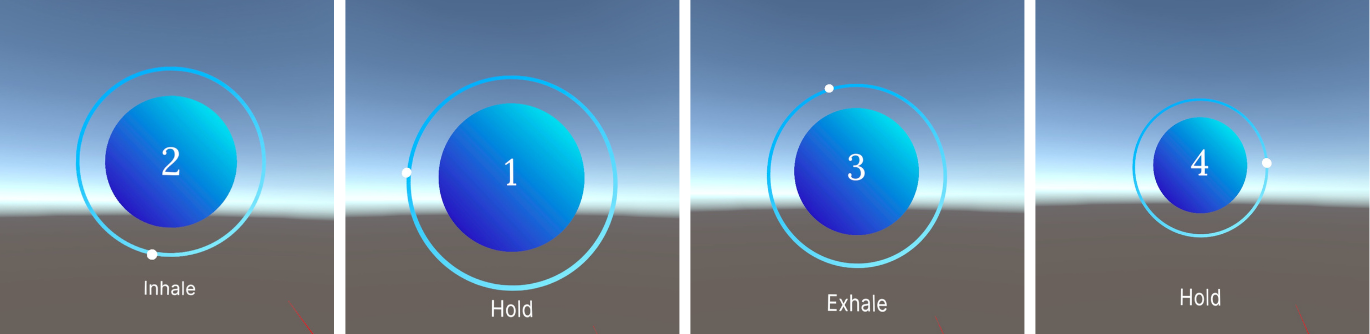}
\caption{Onboarding system of box breathing guidance. The user is asked to inhale, hold, exhale, and hold for 4 seconds each. The bubble expands and contracts following the breath, with a countdown timer for 4 seconds for each stage.}
\label{onboarding}
\end{figure}

\begin{figure}[!t]
    \includegraphics[width=\textwidth]{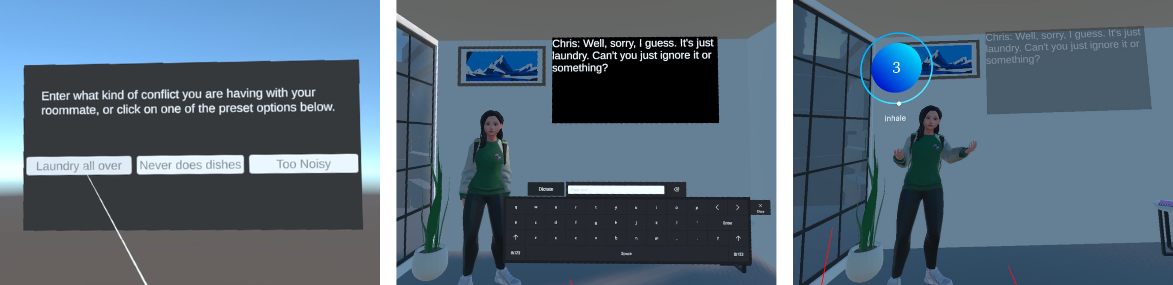}

  \caption{Example of interaction flow in the VR prototype for roommate conflict. The user can select a topic (left) then begin a dialogue with the generated scene (middle). At any point, the user may trigger the breathwork guidance and conduct box breathing while the simulation is paused (right).}
  \label{fig:workflow}
\end{figure}

\textbf{Guidance System Development.} For VR/AR prototypes, the box breathing guidance was informed by prior research to use cool colors \cite{lukic2021physiological}, abstract imagery \cite{tan2023mindful}, and expansion and contraction to guide the breathing pattern \cite{tan2023mindful}. Given past work's recommendations \cite{patibanda2017life}, we also showed users a dedicated onboarding system (without any scenario simulation) for users to familiarize themselves with box breathing before entering the simulation; screenshots of the box breathing guidance system are shown in Figures \ref{onboarding} and \ref{fig:workflow}. Users could access breathing guidance at any point by pressing a button on the Quest 3 controller. For \textbf{\texttt{Text-Guided)}}, its guidance implementation is output just as text instructions in OpenAI Playground (see Figure \ref{fig:ABC}); we prompt GPT to output breathwork at recommended moments of high stress. We ask users their opinions on this LLM-powered guidance instruction versus manually-selected guidance (like in VR/AR) in our interviews.

\subsection{Semi-Structured Interview}
We conducted prototype-driven interviews with 19 participants. Interviews were semi-structured, guided by a list of questions (see Appendix) but allowed to deviate depending on topics participants may have introduced. The interviews took place in-person, lasted approximately 1 hour, and were audio-recorded.

\subsubsection{Participants and Recruitment}
We aimed to recruit participants of diverse backgrounds, so eligibility was only restricted to those 18 years and above and required in-person presence for the study. We recruited participants through putting up physical flyers in the local community, posting on social media, and snowball sampling. Interested respondents were asked to complete an intake form that asked for their basic demographic information of gender identity, age, and racial identity. Interview participants' demographic information is listed in Table \ref{participants}. In total, we interviewed 19 participants, who we met in-person. Participants were compensated with a \$20 Amazon gift card. This study was approved by the appropriate Institutional Review Board (IRB).

\begin{table}
\centering
\caption{Participant demographics for gender, age, and race.}
\begin{tblr}{
  width = \linewidth,
  colspec = {Q[117]Q[181]Q[369]Q[54]Q[204]},
  row{odd} = {Mercury},
  column{4} = {r},
  hline{2} = {-}{},
}
\textbf{Participant} & \textbf{Gender}& \textbf{Race / Ethnicity}& \textbf{Age} & \textbf{Exposure Condition} \\
P1 & Cisgender Man  & Asian  & 21  & Public Speaking Q\&A \\
P2 & Cisgender Woman& Asian  & 24  & Public Speaking Q\&A \\
P3 & Cisgender Woman& White  & 21  & Social Party \\
P4 & Cisgender Man  & Hispanic, Latino, and/or Spanish Origin & 21  & Public Speaking Q\&A \\
P5 & Cisgender Woman& Black and/or African-American  & 19  & Interpersonal Conflict\\
P6& Cisgender Man  & Asian  & 30  & Public Speaking Q\&A \\
P7 & Prefer not to say & Black and/or African-American  & 29  & Interpersonal Conflict\\
P8 & Cisgender Woman& Asian  & 19  & Interpersonal Conflict\\
P9 & Cisgender Man  & Black and/or African-American  & 21  & Interpersonal Conflict\\
P10& Cisgender Man  & Asian, White & 22  & Social Party \\
P11& Cisgender Man  & Asian  & 42  & Social Party \\
P12& Cisgender Woman& Asian  & 24  & Interpersonal Conflict\\
P13& Prefer not to say & White  & 41  & Social Party \\
P14& Cisgender Woman& Asian  & 29  & Social Party \\
P15 & Cisgender Woman& Asian  & 20  & Interpersonal Conflict\\
P16& Cisgender Man  & Hispanic, Latino, and/or Spanish Origin & 22  & Interpersonal Conflict\\
P17& Cisgender Man  & White  & 38  & Public Speaking Q\&A \\
P18& Cisgender Woman& Black and/or African American  & 19  & Interpersonal Conflict\\
P19& Cisgender Woman& Asian  & 23  & Public Speaking Q\&A 
\end{tblr}
\label{participants}
\end{table}

\subsubsection{Exposure Condition.} Participants were asked to report their expected stress and anxiety levels for the three scenarios: (1) answering questions from an audience after their public speaking, (2) attending a crowded social party, and (3) having an interpersonal conflict with a roommate. Participants rated their stress levels ("On a scale of, how stressful or anxious do you anticipate yourself to be when...") for these three situations using the Subjective Units of Distress Scale (SUDS) scale from 0 to 10, which is an established rating system in psychology \cite{kiyimba2020clinical,wolpe1969subjective} (see Appendix for full scale). After participants finished this survey, we recorded participants' highest stress situation and asked participants if they felt comfortable experiencing simulation for it. All participants said yes; we informed participants that at any point they could ask us to change scenarios or stop the interview altogether. For participant safety, we also automatically excluded showing them scenarios for which they rated their anticipated stress 9 or 10, and also provided all participants with the local county mental health hotline number for extra precaution. 

\begin{figure}[t!]
\includegraphics[width=0.7\textwidth]{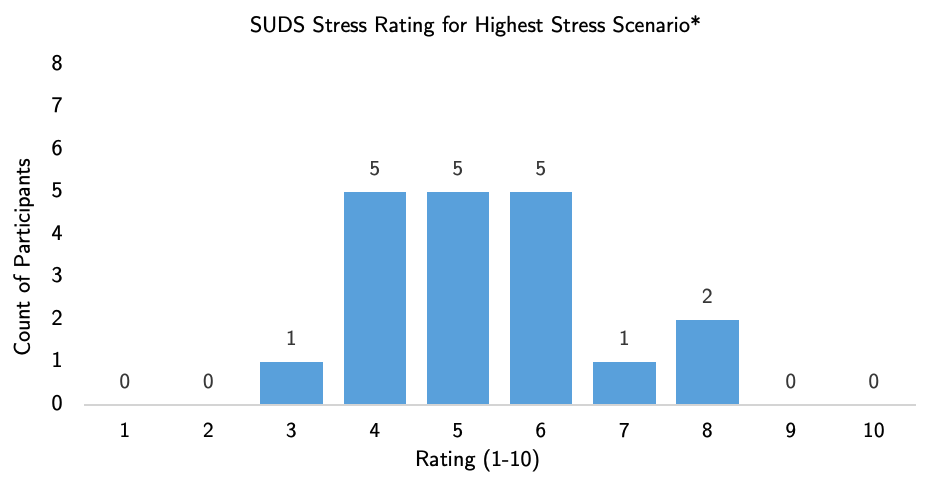}
\caption{SUDS score rating among participants for their exposure condition}
\label{suds-rating}
\end{figure}

\subsubsection{Interview Protocol.} We showed the eight prototypes on the participant's highest stress topic given their initial survey, \textcolor{black}{in the progressive order listed in Section 3.2 (\textbf{\texttt{VR-Static}}, then \textbf{\texttt{VR-Interactive}}, then \textbf{\texttt{VR-Guided}}, ...). We followed this order so that participants had a progressive familiarity and learning of additional features, from no interaction, to added interaction, and finally added guidance features.} In addition, we also showed the "onboarding" guide to the breathing guidance system before the participant saw \textbf{\texttt{VR-Guided}} or \textbf{\texttt{AR-Guided}} to help users engage with the exercise \cite{patibanda2017life}; in the onboarding guide, the participant saw only the breathing orb and no other simulation (see Figure \ref{onboarding}), during which the participant could take as long as they needed to practice the box breathing pattern before proceeding. For each prototype, we provided an introduction blurb for participants to imagine. For example, before showing participants the VR prototypes for interpersonal conflict with a roommate, we stated: "\textit{Imagine that you recently moved in with a new roommate. Things have been tense, though, and you plan to confront your roommate at home later today after work about an ongoing source of disagreement between you two. Nervous about how this conversation will go, you put on a virtual reality simulation to practice what you are going to say to them. In this scene, you’ll be met with your roommate who will speak with you about this ongoing conflict. You’ll be able to choose the topic of the conflict.}" For the public speaking and interpersonal conflict scenes, participants could choose their own topic of choice for their speaking topic and conflict reason, respectively, or choose from pre-sets as shown in Figure \ref{fig:workflow}.

We aimed to understand people's needs, whether this technology met those needs, and understand the design opportunities to improve self-care technology. The interview protocol consisted of showing each of the eight prototypes and asking participants questions revolving around four main concepts: (1) \textbf{initial impressions}, (2) \textbf{usefulness} ("\textit{To what extent do you feel this is useful?}", "\textit{How does practicing stress relief in this way compare to your current practices?}") (3) \textbf{realism and natural interaction} ("\textit{To what extent did the interactions feel natural or unnatural?}", "\textit{What aspects felt realistic or unrealistic?}") and (4) \textbf{general feedback} ("\textit{To what extent did you like this design?}", "\textit{What additional features would be useful?}"). For the prototypes involving the breathing guidance, we additionally asked participants what their thoughts were on using guidance or, if they didn't use guidance, their reasoning for not using the breathing guidance system. Lastly, we asked participants about their experience as a whole with the simulation as well as their general thoughts on applying simulation to their everyday stress relief practices. Given the semi-structured nature of the interviews, our research team's interviewer would also ask follow-ups, clarifications, and ask participants to expand on thoughts they expressed in an evolving nature during the course of the interview. The full list of interview questions are in our appendix.

\subsection{Data Analysis}
After completing all interviews, we anonymized, transcribed, and coded all 19 interview transcripts. Based on Braun and Clarke's method of thematic analysis \cite{braun2012thematic}, our iterative analytic cycle consisted of: (1) recording and transcribing interviews (2) coding transcripts (3) amalgamating codes (4) discussing codes (5) highlighting themes (6) writing and revising memos. Our interviews and analysis continued until saturation, where new interviews gave no new insights. We first generated the lowest-level, granular codes through each transcript, then organized these codes into axial-codes. Afterwards, we arranged these findings into five high-level themes presented in this paper. 
\section{Results}
We organize our findings around five high-level themes identified through our interview analysis, including people's current experiences of stress (4.1), overall perspectives and experiences using simulation for self-care (4.2), and findings around our three high-level dimensions for designing simulation for self-care: modality (4.3), interactivity (4.4), and guidance (4.5). 

\subsection{Current Experiences of Everyday Stress}
Participants experienced a range of different stressors in their everyday lives, most of which were in the themes of social anxiety and work. 12 out of 19 participants described their everyday experience of stress as related to social anxiety, including being in social settings with people they don't know (P3, P12), being in crowded spaces (P10, P12), giving presentations (P2, P14, P15), and talking on the phone (P19). Other sources of stress include taking exams, meeting deadlines, and generally having a lot of work or personal tasks to complete.

\subsubsection{Current methods to handle stress.}

In addition to asking participants what types of everyday stress they experience, we also asked them how they dealt with stress. While participants did mention a range of coping mechanisms, such as relying on social connections (P11, P14) and problem-solving (P3, P10, P14, P15), the most common theme across participants was using distraction.

\begin{quote}
    "\textit{talking to people about lighthearted things...so I can \textbf{distract} myself}" (P5)

    "\textit{[I use] messaging applications...talk about random stuff and \textbf{distract} myself from any conflicts going on.}" (P16)
    
    "\textit{\textbf{distract} myself a bit, browsing social media}" (P1)
    
    "\textit{I usually \textbf{distract} myself}" (P12)
    
    "\textit{Not a technique, but I do \textbf{distract} with hobbies that are not related}" (P16)
    
\end{quote} 

Regarding strategies targeted towards mitigating their own stress, many participants had tried habits like journaling (P14, P17), meditation and meditation apps (P3, P12, P14), and breathing techniques (P1, P2, P4, P5, P11); however, almost all did not sustain these practices over time with the exception of regular exercise, which was cited as effective at keeping chronic stress low (P4, P7, P8, P9, P13). Participants' reasons for not continuing use of self-care techniques during stressful situations ranged from finding some techniques \textbf{difficult} ("\textit{Journaling was a thing that my therapist said, but I don't have patience for it…I wish it would be more straightforward somehow}" (P17), "\textit{I have briefly tried meditation apps... I just kept not focusing. So [it] didn't really go that well.}" (P8)), a \textbf{lack of enjoyment} ("\textit{I downloaded a few apps, but again, I find it annoying.}" (P17)), \textbf{accessibility issues} such as cost ("\textit{I used Headspace [a meditation app], but I don't really do it anymore because I didn't want to pay the subscription}" (P14)), and \textbf{start-up work} (“\textit{it's a lot of work that you need to get in touch with [a] therapist, and you need to experiment with them to see if you work well together.}” (P19)). Others had no particular strategies at all to dealing with stress in their everyday, such as P19 who said they "\textit{just wait till [the stress] is gone}". 

\subsubsection{Perceived ineffectiveness of current stress reduction habits.}

When describing their current experiences of everyday stress, participants generally did not feel that they had good techniques for stress relief (e.g. "\textit{I don't think I've had a ton of success}" (P3)). In particular, \textbf{participants felt they lacked self-sufficient skills}. For example, P16 said that their tendency to use distraction to deal with stress was "\textit{evasive and destructive, which has worked enough, but I feel like it relies on something else... And I do think I don't have the techniques or knowledge to just by myself meditate or calm down}" (P16). When speaking about how she copes with making calls to customer service, a source of stress for her, P19 expressed, "\textit{If I need to make the call by myself, I think I don’t have a really good way to do that}” and this caused even more stress around not having a good way to deal with that stress: "\textit{I think I feel stressed by not having a good way to cope with [the stress]}" (P19). 

Additionally, heightened emotions was cited as a barrier for handling acutely distressing or stressful situations, such as interpersonal conflict. For these participants, calming their emotions was often seen as a necessity for them to them properly destress or self-soothe: "\textit{I have to calm down a little bit. My reaction is a lot of anger. I clench my teeth, then I start sweating [and] my eyes become a little hazy...I do nothing until I’ve calmed down}" (P6). Another participant expressed, "\textit{Anger really makes [it] hard to get out of [stressful] situations... it might be good for me to practice how to control my feelings by practicing breathing"} (P11).

\subsection{Overall Perspectives on Using Simulation for Self-Care}

After understanding participants' current experiences of stress, we proceeded to gather their thoughts on our prototypes and overall perspectives on the role that social simulation may take in their self-care.

Overall, participants responded very positively to some or all of the prototypes shown as well as the general idea of using virtual simulation to practice stress relief. \textbf{17 out of 19 participants said they would use one or more of the shown prototypes} (VR, AR, or text-based) in their current lives if they had access to the technology. Participants felt using simulation would \textbf{improve the outcomes} of situations ("\textit{it would easily complement what I would normally do in reality, and just make all these situations work out better for me}" (P9)), help \textbf{explore different responses} ("\textit{Just being able to articulate it beforehand and think, "okay, maybe I should try saying this instead". I think it would be helpful for me} (P18)), make them \textbf{feel more prepared} ("\textit{actually having this scenario completely play out before it plays out} (P5)") and \textbf{mitigate negative emotions} ("\textit{you can use it to mitigate some anger, stress}" (P9)). People even re-imagined the use of this technology, such as P16 who mentioned that they feel it would be useful in professional therapeutic settings rather than just for personal self-care, saying, "\textit{If a therapist had this available during a session, it could be quite useful to test-drive situations or problems that you're addressing during therapy.}".

\subsubsection{Benefit of realistic self-care practice}
Participants felt that the main benefit of social simulation was to practice stress relief techniques they could use "in-the-moment" of real-world scenarios. Doing so within a realistic environmental setting was praised by participants and seen as a step up from their current ways of coping with stress, as P16 notes, "\textit{The closest that I'd imagine doing before I came into the study would have been practicing something in my head or in front of the mirror, which compared to this, is way less evocative and doesn't put you into the same mentality. So it wouldn't be as useful or as effective as something like this}". P5 also appreciated the practice "\textit{of tackling the stress by just being in the situation itself}" as it helped them "\textit{mitigate the fear}" when approaching an interpersonal conflict. In contrast to retroactively reflecting, the simulation helped surface emotions in-the-moment of stress, which could result in a more accurate and reflective picture of one's emotional state. For example, one participant noted that  "\textit{I feel like this one, I can do it in the middle of whatever I'm experiencing. I definitely feel more in control because other techniques, you just reflect on it afterwards. With this one, I feel more like reflecting in that moment, trying to observe my reaction. In that sense, it was more helpful}" (P14). Moreover, participants noted that doing the breathing guidance served as a reminder to calm themselves "\textit{in the middle of a situation}" when they expected to be too emotionally elevated and "\textit{may not really come up with that idea [to] breathe at this moment}" (P11). Our simulation was also seen as fostering better transferability of these skills to real-life situations compared to interventions that remove individuals from the context of stress. Participants enjoyed the practical aspect of practicing deep breathing while "in" the stress-inducing environment, versus other technological interventions for self-care that either have no environmental immersion \cite{lahtinen2023effects} or use abstract or "escapist" (e.g. forest, beach) visuals \cite{roo2017inner, ng2023virtual}:

\begin{quote}
    "\textit{There's two types of things, right? There's one thing that you do on a daily basis just so that you don't get triggered as much...I feel like meditation falls in one of those, whereas deep breathing or stress balls or whatever people do -- those fall into in-the-moment stress. And I feel like this is great for the second part...I could see myself using this, where I know a situation is probably going to trigger some stress for me. I would like to take all of those layers out as much as I can [by] using something like this before I actually enter into that situation.}" (P12)

\end{quote}


\subsubsection{Building Confidence and Preparedness}

Participants also found the non-guidance versions (e.g. \textbf{\texttt{VR-Interactive}}, \textbf{\texttt{AR-Interactive}}, \textbf{\texttt{Text-Interactive}}) useful for simply preparedness and confidence having already had (virtual) experience with a situation beforehand. P18 articulated that this practice "\textit{has the use of getting over the initial nervousness of confrontation}" and helped "\textit{[get] over anxiety and trepidation}". P15 mentioned feeling more prepared after just using VR-Interactive without any mental health guidance: "\textit{I would feel more prepared to actually confront them. The [avatar] showed me some attitude and I was kind of caught off-guard. If my roommate actually did that, I would have been prepared}". The technology also helped de-sensitize to anxious situations as one participant notes regarding her experience coping with social anxiety through avoidance: "\textit{you start avoiding places that make you uncomfortable...but some stress you can't avoid}" and described using the simulation as a "\textit{choose-your-own-adventure}" tool to "\textit{practice stepping out of your comfort zone}" and to help her "\textit{reintroduce to social situations}" (P13).

\subsection{Modality: Benefits and Potential Risks of Immersion and Realism}
We summarize below how participants responded to different modalities and allowances for immersion in our prototypes. Overall, participants expressed a desire for increased realism in the prototypes but also raised concerns about emotional safety.

\subsubsection{Preference for Augmented Reality.}
\textcolor{black}{Rather than expressing a universal favorite among the modalities shown in the study,} many participants saw value in multiple designs depending on context:

\begin{quote}
"\textit{If I'm out in a public scenario, and putting on the VR headset is a little much...then I started typing away. I think that's the best use for [the text-based chatbot]. For me, the AR one [if] I'm in the area in which I'm going to have the talk... The VR one [if] I'm inside, I'm alone.}" (P9)\vspace{1mm}

"\textit{Using the VR is like the initial part, just to only be able to focus on the conversation. The AR is the next step where you're in the space where the issue is...so [first] being able to practice in a controlled environment, which is the VR, then move to the AR.}" (P18)

\end{quote}

One notable consensus among participants, \textcolor{black}{though, was that they were most likely to use AR if they had access to \textbf{the real physical locations where anticipated stressors might occur}; for example, participants} said that they'd "\textit{use AR when it is close to where I will actually give a speech}" (P19) such as in a classroom, and practice stress-relief for a roommate conflict "\textit{in my actual room...maybe I can breathe and walk around while I breathe}" (P9). Indeed, "\textit{a lot of the confidence comes from just being comfortable in your space}" (P12) and practicing situations in the expected environment was already what some participants said they would do now if preparing for something like public speaking. This would also allow for people to practice their own \textbf{physicality}. For example, P9 felt AR would "\textit{encourage me to move around a little bit, do hand gestures}" and, more than the dialogue, P17 saw the simulation as more useful for his "\textit{physicality, you know moving around...engaging with people, which approaches I would do. I could train that}".

Experiencing AR also opened up suggestions of different mental health guidance options according to people's personal needs, such as navigating their surroundings to reduce stress. For example, P11 felt he would use AR to practice in his real home being able to retreat to a different room to do deep breathing if overwhelmed at a social party. Similarly, people could also practice doing guidance rooted in visual stimuli in the environment as P12 mentions: "\textit{I know a few people who do struggle with social anxiety or confrontation. I remember somebody had mentioned to me that...there's a point in their room that they look at and that helps them just go into zone of 'I need to breathe three times'. [AR] could translate to that...you can just look at the same spot, and it should prompt the subconscious breathing}".

\subsubsection{Environmental Realism}
Although past work in social simulation has focused on agent interactions to simulate a "real-world" setting \cite{park2022social,ren2014agent}, we found that \textbf{environmental factors are critical in participants' experience of social simulation}; immersive simulation goes beyond just accurate interactions or behaviors. In particular, ambient sound and visual immersion were noticed and desired by participants. For example, multiple participants felt that the background sound in our prototypes was key for their feelings of presence ("\textit{listening to the sound [of the party] made me anxious}" (P11), "\textit{the sound was very important [for immersion]}" (P17)) and wanted even further sound immersion, such as emphasizing particular sounds ("\textit{In real life, when you’re stressed or anxious, certain sounds are emphasized, and it adds to the feeling of stress}" (P17)).

We saw the importance of environmental factors for realism when comparing users' reactions between AR/VR to the text-based simulations, which participants described as "\textit{artificial}" (P11) and "\textit{more like a tech simulation}" (P10) despite GPT prompting and dialogue style being consistent across VR, AR, and text-based interactions. P16 pointed out the stark difference between the visual experience of AR/VR versus text: "\textit{I'm comparing it now to what I was experiencing during the AR/VR experience...it feels a little more clinical, removed. Obviously it's the same engine and it's the same conversation model, but the act of either being in AR or VR space and having, albeit rudimentary, a representation of another person does help the believability and the immersion}". The disconnect participants felt in text-based simulations despite the same dialogue prompting as in VR/AR showed the importance of creating rich multisensory environments. 

Participants consistently expressed a need for customizable avatars and environments to better reflect real-life scenarios: "\textit{This scenario could be more personalized. For instance, if I could say, 'I'm attending this party tonight and meeting these specific people, and I'm nervous about it,' then maybe the conversations could reflect that}" (P14). Participants desired for simulations that not only felt plausible but were specifically relevant to their own lives. Even smaller physical differences, like positioning of avatars, could break immersion. For example, P12 pointed out that the roommate conflict simulation felt unnatural because the avatar stood facing across the user. She commented that she would not imagine this positioning for a conflict situation and desired the ability to customize this, stating, "\textit{'do you want to be standing up or sitting down?' -- maybe just as simple as that. The room that you have right now, I would choose to sit on the couch}". Generally, we found that participants desired modification features for avatars and environmental details to match their real-world expectations.

\subsubsection{Risks in Realistic Simulation of Stress}
While participants found our prototypes to generally be immersive, \textbf{they wanted the simulation to produce higher stress levels} to more closely emulate real life. Participants saw this as giving "\textit{insights [on] how to handle or how I would feel...Basically, it's a way to anticipate my stress or my anxiety in a real setting}" (P17) and avoiding getting false security since "\textit{[if you] go to practice with this, and then you go for a real environment with real people -- maybe they [will not] be so gentle with you}" (P7). P10 echoed this and surfaced a potential limitation with using GPT for interactive dialogue, saying, "\textit{I went in with the impression that no matter what I responded with, there would be people bringing up other new topics and things that I could respond to. In real-life conversation you're also obligated to keep the conversation flowing}". In fact, many participants correctly identified our dialogue engine as GPT-generated, and commented on GPT being unrealistically polite. Indeed, several participants chose not to use the available guidance during \textbf{\texttt{VR-Guided}}, citing a lack of stress. Although some felt guidance would be useful, they did not need it during the simulation: "\textit{If the roommate would have persisted a little more, I would have probably [selected the breathing guidance]...maybe I just wasn't stressed enough in this situation}" (P12). 

However, the risk of simulating high levels of stress is \textbf{potential risk of trauma}. For example, when P19 was asked whether she would like the simulation to be more realistic, she expressed hesitation and stated, "\textit{I sometimes feel like I would be traumatized when I'm having conflict with somebody. I just want it to happen once. And I think practice will help to reduce the stress, but I kind of feel like maybe I'll get traumatized}". Another participant noted, "\textit{If somebody has a very deep fear of public speaking, this might be too realistic...confronting the fear head on, depending on the person, might help or might not}" (P5). Some participants felt lacking some realism created a safer environment for practice. P8 expressed that greater realism "\textit{kind of scares me}" and P5 appreciated the current balance as helpful but safe: "\textit{I do like that it's not completely realistic. It helps me remember that it's just a simulation and that it's more for mental health}". The current level of (or lack of) realism allowed users to feel safer in verbal expression and more willing to engage in difficult scenarios. This allowed participants to feel it was "\textit{okay to fail in the conversation}" (P14) and "\textit{it’s a safe environment... I know it's virtual [so] I am more relaxed...it will not have a repercussion}" (P17). 

\subsubsection{Value in Low-Immersion Designs.}
Despite a preference for immersive features, participants found value in low-immersion, text-based prototypes due to their accessibility and ease of use. The ease with which users could access text-based simulations made them a practical alternative for stress-relief practice in everyday settings. Participants mentioned that accessing chatbots “\textit{might be more accessible for most users}" (P15), and “\textit{in your everyday, I think this is way easier to pull up on your computer or your phone...like VR is something that most people either don't have or don't have on them always}" (P8). As a result, while environmental immersion was an important factor to people finding simulation helpful for practicing stress relief, participants still valued non-immersive but accessible systems.
\subsection{Interactivity}
Interactivity through dialogue was preferred across the board among participants. When asked, all participants explicitly said they preferred \textbf{\texttt{VR-Interactive}} and \textbf{\texttt{AR-Interactive}} over the static counterparts due to the dialogue. However, despite people preferring interactivity, non-interactive designs were still seen as more valuable than current means of practicing stress relief. 

\subsubsection{Interactive Dialogue}
Having interactive, realistic dialogue was key for participants in evoking accurate reactions and emotions. The interactive nature evoked genuine emotional responses, which participants felt surprised but pleased by, that fostered deeper engagement and self-realizations. For example, P11 said "\textit{I didn't really think the headset [could] make me very stressful...I was very surprised}" and the dialogue even evoked hesitancy akin to how participants would feel in real life:

\begin{quote}
"\textit{Even though it's a simulation, you could hear [my] stuttering and the hesitation. I was trying to still not go in with too much anger or irritation, but then also trying to be as direct with the problem.}" (P18)\vspace{1mm}

"\textit{Even now, I was stumbling over my words and I wasn't really sure what I would say. So this is a perfect scenario, where I could think of how to approach the situation delicately.}" (P5)
\end{quote}

Participants commented on enjoying how they were prompted to give realistic responses. P12, who engaged in an interpersonal conflict situation, stated that they enjoyed how "\textit{it was not like they suggest the solution upfront because I feel like most of the times, it doesn't go that same way}" and P10 felt questions received during the public speaking scenario were "\textit{in-depth}" and "\textit{elicit a very thoughtful response}". 

Beyond just realistic dialogue, participants emphasized that they desired more nonverbal cues such as facial expressions and body language due to its relevance in real life. P14 pointed out that "\textit{nonverbal behavior is what's causing a lot of the anxieties as well}" and P19 also felt so regarding public speaking, stating, "\textit{I would prefer [the avatars to] have more facial expression when I answer some questions...some [audience members] will make some faces that make me really nervous...it also adds more stress to myself, but I would say it will make the scenario more realistic}".

All 19 participants favored dictation using our speech-to-text feature over typing on the VR keyboard due to ease of use and its similarity to real life interactions. However, since our designs did not implement text-to-speech, nearly all participants expressed wanting audible voices from avatars. As P19 notes, avatars speaking out loud rather than our design of showing dialogue in avatars' text boxes would help give "\textit{a shorter time to process the information}" to create "\textit{more realistic interactions}" akin to real life. 

\subsubsection{Movement and Interaction with Environment}
Movement and interaction with environment were interactions that participants expressed wanting more of. As we reviewed previously, people enjoyed the thought of AR due to their closeness to the real environment they expected to encounter stress. Some of this desire came from expecting physical movement to play a significant role when feeling discomfort. People therefore felt that the fact they were stationary in our VR/AR prototypes was limiting. For example, P13 noted that their natural response to feeling socially anxious would be to move around, "\textit{my impulse would be to walk to a new location within the space. A lot of people in a social situation, if they see a snack table will be like, "look, snacks!" and you run over there}." The ability for greater levels of interaction with the environment could help users also simulate different ways of handling stress, such as finding more comfortable spots in a room or occupying themselves with other objects. Additionally, allowing interaction with the avatars based on the simulation's environmental factors could help people practice icebreakers and conversational skills in a more natural way, as P13 points out about how socially anxious people may choose to converse when feeling nervous in social settings: "\textit{it's the cat, or these plants are nice. You look for some shiny objects and start a conversation based on artwork in the room or something}".

\subsubsection{Value in Non-Interactive Simulation.}
Users enjoyed the realism and interactive aspects of the prototypes, as reviewed above. However, interestingly we found that participants still perceived value even in the prototypes without any interaction (\textbf{\texttt{VR-Static}} and \textbf{\texttt{AR-Static}}) as they were still perceived as more helpful than participants' current means. For some, this was due to the fact that they added any kind of visual to an already existing simulation people currently do in their heads to prepare for difficult situations ("mirror the act of pulling it up in your head" (P16)). Similarly, regarding \textbf{\texttt{VR-Static}}, P12 felt, "\textit{I'm comparing it to practicing in the mirror...it's the same fundamentals. You're trying to get used to seeing things that you're not normally used to saying...it is good}". The lack of realistic graphics was also not seen as necessary to still gain value from the technology for most participants: "\textit{I don't think it's a problem that it's low poly or not necessarily realistic graphics. The act of being in this space and having a roommate character in front of you makes it into the psychology of being in that situation}" (P16).

\subsection{Guidance Needs}
Although participants still found the prototypes without mental health guidance useful, breathing guidance in VR-Guided and AR-Guided prototypes yielded an opportunity for \textbf{emotional regulation} and \textbf{more productive responses}. 

Breathwork helped participants calm themselves and self-reflect on their emotional state. For instance, participants noted breathwork "\textit{definitely made me calm down}" (P14) and helped "\textit{mitigate some anger, stress}" (P9). Participants usually triggered breathwork when dialogue became more uncomfortable, such as a roommate becoming defensive or receiving a pressing question from an audience member. In doing so, participants were able to be "\textit{more aware of the situation and gives me some time to actually watch what kind of emotions that I have in myself}" (P14) and that breathing during a more stressful encounter "\textit{actually did a good job in helping manage mostly anger}" (P9).

Additionally, many participants also mentioned its usefulness for helping them respond more productively. For instance, participants stated "\textit{with that breathing exercise, it will help me to be calm and think about the answer}" (P17) and "\textit{being able to articulate it beforehand with a response and think, "okay, maybe I should try saying this instead}" (P18). Pausing during stressful moments gave a moment to "\textit{recenter and get back to what I practiced}" (P18) as a healthier response to unexpected stress arising. Additionally, given participants' current methods of stress relief revolving around avoidance and distraction (see Section 4.1), it was particularly notable that participants felt that practicing breathing would help them to not immediately leave situations in unproductive ways. For example, P9 commented "\textit{if I didn't press the breathing bubble, I would have just been like, 'okay, it's whatever' and then left when in reality, that's not going to resolve a situation. I think taking that extra time to breathe there was better}".

\subsubsection{Transferable Skills to the Real-World}
While some participants felt the practice was good for habit-building for applying to real-life (e.g. "\textit{you can make it a practice...do it subconsciously so that it calms you down}" (P12)), other participants expressed that deep breathing was not necessarily a transferable practice due to its \textbf{impracticality during real-world situations}. As a result, we discovered that \textbf{transferability, practicality, and even social acceptance} were key dimensions that affect whether mental health guidance is perceived as useful. For instance, participants expressed hesitation on doing box breathing in real life:

\begin{quote}
    "\textit{I do kind of worry that in real life people are not gonna wait for me to breathe and breathe out.}" (P14) \vspace{1mm}

    "\textit{People will think: why the long pause?}" (P6) \vspace{1mm}
        
    "\textit{I can't stop this conversation anytime without worrying about [the avatar's] feeling. So I can't just roughly [and] abruptly stop talking with him without answering his question.}" (P11)
\end{quote}

As a result, people wanted and suggested different types of guidance that they thought were more relevant to their specific lives. For example, P19 suggested that her real-life calming mechanism during public speaking was seeing encouraging facial expressions from the audience and thus recommended simulating this. P18 mentioned techniques that felt easier to do in the middle of a conversation, such as, "\textit{closing your eyes and counting to a certain number and then gather your thoughts and recenter yourself}" (P18). Other participants simply just wanted options, such as "\textit{not only the breathing, but maybe like a couple other options I could click for how to relieve the stress}" (P9), or what they felt would be more relevant to the context: "\textit{The assistant was being sarcastic and the guidance said to practice box breathing, but I don't think I needed box breathing for that. Maybe if they were being a little bit more aggressive, I would need box breathing}" (P15). 

\subsubsection{Manual Guidance: Empowering users' agency and self-reliance}

We also investigated participants’ preferences between GPT-timed breathing guidance and self-triggered guidance. One reason people preferred manual control was their confidence in recognizing their own stress. For example, one participant commented on how she understood her needs when the moment arose: "\textit{It was kind of vague in the beginning whether I should turn [breathwork] on or not. But when the moment came that I felt, "oh, maybe this is a good time to take a break"...I just made more [of a] conscious decision on when I need[ed] that}" (P14). Participants also felt manual control better reflected real-life decision-making, encouraged habit-building, and facilitated learning. As P9 explained, "\textit{It's more realistic and applicable if I control it, because in real life I would be controlling when I do something like that}" and P16 echoed, "\textit{Realistically, you would be the one trying to execute your own calming techniques...I think you yourself are the best arbitrator of when something's stressful}."

More explicit concerns over \textbf{intrusiveness} also arose. P9 commented, "\textit{I wouldn't like it if...the simulation made me do the breathing thing, because maybe I didn't feel like I needed it. Or I felt like I could handle the situation without doing the breathing}" while P7 similarly felt "\textit{forcing someone in middle of the conversation might make things much worse...what if I don't want to take a break?}". These automated recommendations could even cause people to feel self-conscious; P14 encountered a suggestion for breathing guidance during \textbf{\texttt{Text-Guided}} and stated, "\textit{I felt like I wasn't feeling stressed, but I think I got direction that I'm stressed...Like, am I stressed? When I'm not stressed, do I look stressed? I don't really like being told that I look stressed}". 

Interestingly, participants also expressed a general distrust in the accuracy and universality of AI-driven mental health guidance. Even participants who found GPT's recommendations accurate during the study remained skeptical about future scenarios. For example, P9, who found the breathing suggestion useful, still had reservations: "\textit{I actually thought the system did pretty good on this one, I thought the situation was getting a little stressful...but I don't know if it will do that every time. But for this time, it did good}".

\subsubsection{Automated Guidance: Enhancing Self-Awareness}
On the other hand, automated guidance helped users reflect on their emotional state and thus could lead to gaining self-awareness. While some participants expressed concerns about system-determined guidance, others appreciated its value in moments when stress might go unnoticed. P19 shared that she preferred LLM-powered guidance suggestions due to a lack of anticipation of her own stress, stating, "\textit{When I was answering some of the questions previously, I did not feel like I'm stressed until I was stuck. Then I started to feel stressed, and that's something that I did not anticipate}". For these participants, automated guidance also was more intuitive to include in a training system such as ours in order to develop further self-awareness skills and allow users to consciously notice moments of stress they may have otherwise missed.

Overall, we found that users feel automated guidance can play a supportive role in helping users develop better self-awareness and emotional regulation skills. However, serious concerns over accuracy and agency also emerged. Perhaps due to the distrust of LLM-powered recommendations and its universality, multiple participants surprisingly suggested using \textbf{biosignals} to guide automatic recommendations instead. In subsequent interviews, when probed with the idea, all participants reacted positively to this idea.  For example, P7 commented that "I prefer to take charge. But I guess for such a system, tracking your heart or something for it to be accurate? Or maybe your voice tone." and P16, another participant who suggested this idea, said, "I like the idea of monitoring your vitals and signs of stress, maybe especially in a VR/AR setting. Having like a non-intrusive pop-up or gentle reminder" when the user showed a signal like high heart rate. Integrating biosignal monitoring could also provide an alternative route to an accurate and personalized approach that still facilitates practicing self-awareness around stress management.

\section{Discussion}
This study explored user perceptions of applying social simulation technologies for practicing self-care techniques for their everyday situations. Through prototype-driven interviews with 19 participants using eight built prototypes, we gathered valuable insights into the benefits, risks and challenges, and future design implications for applying these technologies to self-care. 

\subsection{Designing Simulation for Everyday Self-Care}

As Nunes et al. note in a review of HCI technology for self-care, the concept of technological support for people's mental health should be expanded to center on people's everyday life experiences rather than just patients of chronic conditions \cite{nunes2015self}. Our study contributes to the field of HCI technology for everyday self-care and explored what design needs exist in this relatively unexplored space. 

\textbf{Modality.} One of the most significant findings was the clear preference for AR among our prototypes. Participants envisioned using AR in real-world environments to overlay both practice scenarios and mental health guidance, allowing for the direct application of coping strategies in the settings they encounter in daily life. In the context of everyday stress, where individuals are frequently exposed to environments that trigger emotional or mental distress, our interview findings suggest that AR lowers barriers in transferring skills to real-world settings and individuals prefer to practice coping techniques in spaces that closely resemble real-world settings. While high levels of immersion in VR/AR can foster engagement and be beneficial for simulating scenarios that aren't immediately accessible \textcolor{black}{\cite{shin2019does, bowman2007virtual}}, its separation from real-world contexts may limit the practical application of skills learned in the virtual space. Text-based prototypes, unsurprisingly, were considered the least immersive given absence of visual and auditory feedback. However, many participants said they would use text-based systems because of its availability throughout the day, suggesting that even non-immersive tools have a valuable role in the self-care space.

\textbf{Interaction.} Participants clearly favored features like dialogue and environmental interaction, as these allowed them to practice communication skills in a system-responsive manner. This interactivity provided more opportunities to experience realistic stress and practice self-soothing techniques in response. Notably, even non-interactive prototypes like VR-Static and AR-Static were considered more helpful than participants' current methods of stress relief, indicating that while interactivity is desired, it is not essential for perceived usefulness.

Participants' reactions to our prototypes echoed past findings that LLMs are capable of generating \textcolor{black}{natural, realistic dialogue \cite{gray2024increasing, kim2024understanding, duan2023botchat} that participants found comparable  to their expectations} of real-world interactions. In the future, given the success of speech-to-text in our study, LLMs could be used to also dynamically design environments rather than just provide dialogue as it did in our study. Given the accuracy of our speech-to-text engine using WhisperAI, it is plausible that users could verbally describe anticipated situations, and the LLM could then generate corresponding dialogue, character personas, physical avatar appearances, and environmental features to create a VR or AR scene for the user to practice with. Since participants consistently expressed a desire for customization features to adjust simulations to reflect their personal environments and stressors, using LLMs for designing more aspects of the simulation could facilitate this capability while reducing the time, effort, and technical skill needed on the user's part.

\textbf{Mental Health Guidance.} As seen through our interviews in how people responded to box breathing and their original approaches to self-care in the initial parts of our interview, there is not a one-size-fits-all technique for in-the-moment stress relief. Overall, designs should opt for flexibility or adaptability in guidance. In terms of our specific system, though, participants found the guided interventions useful during moments of heightened stress. However, an important theme that emerged was the transferability of such techniques to real-world interactions. Participants expressed concerns about the practicality of taking a moment to themselves during a heated conversation or stressful social event. Although one method of addressing these concerns is to offer techniques that are more easily transferable, this feedback might also suggest a greater opportunity to also teach people how to advocate for their own self-care needs in social settings and/or learn to transition into self-soothing skills during social settings. For example, instead of social simulation pausing the scene when the breathing guidance was triggered as in our study's designs, social simulation could also continue the interactions while guidance appeared and even simulate realistic ways that people may respond during these moments (e.g. facial expressions, asking a question again if the user does not respond in a timely way). By doing so, users can practice interaction necessary to pause, transition into self-soothing techniques, or step away from a conversation to manage stress. While our prototypes focused on the concept of individuals learning these skills to incorporate into their own daily lives, our interviews also surfaced the significant role that social dynamics play in the application of self-care strategies and even the interpersonal consequences of doing so.

\subsection{Ensuring Safety and Agency through Customization}
Although our simulation is designed as a safe environment to help users practice difficult situations, simulation can also carry risk in itself. The most apparent risk that emerged through our interviews was a tension between people wanting more realism, but also having concern over the potential of discomfort and trauma. Balancing realism and virtual imagery is crucial, as it impacts both user comfort and the risk of trauma. As one participant had noted in our study, stressful situations like interpersonal conflict can be difficult enough to deal with already just once in the real-world. 

Our recommendation for giving users more control, minimizing surprises, and preventing stress responses beyond users' boundaries is to implement several \textbf{customization features} in the simulation. Customization features were mainly suggested by participants as contributing to realism, but that same control can ensure safety for users' boundaries as well. Customization of avatar and environmental features, for example, can be made more real or artificial according to users' preferences. Since sound was a key immersive element as well, simple features like volume control or the option to mute should be available at anytime during simulation in order to reduce auditory immersion if needed. Additionally, a theme that emerged from our findings was \textit{tailoring of difficulty level}, given that some participants wanted more challenging situations to practice with while others had concerns over trauma. Adjustments in mental healthcare to gradually lead up on more challenging situations is well-established in traditional therapeutic settings already, such as "gradual exposure" in Cognitive Behavioral Therapy \cite{cohen2012trauma} and traditional Exposure Therapy \cite{craske2014maximizing}. Challenge in dialogue, such as a roommate being more defensive or combative in our simulation, may be tailored to GPT, and further work may need to be done to find prompting techniques that generate appropriately easy versus difficult output that is appropriate and consistent for users. As a result, we strongly recommend simulation for self-care provide control (and thus more safety) over specific features that may affect one's stress levels (e.g. amount of people at a party, volume levels) as well as difficulty when prompting LLM-powered dialogue. Overall, we find customization not only gives users agency, but also mitigate risks as users navigate their own boundaries. 

As one participant suggested in our study, social simulation has potential to go beyond personal care and could be deployed in a clinical or therapeutic setting for supervised observance. Doing so may help therapists or clinicians see first-hand how their clients react in situations they encounter in life outside therapy, suggest coping techniques, and allow clients to reflect more accurately on feelings during situations they bring up in therapy. Although our work was originally built with a user's own individual, non-supervised learning and practice in mind, it is plausible that learning unsupervised may be best for people who already have personal understanding of their stress \cite{nunes2015self}. Different autonomy levels may be required for different "treatments" -- while some care decisions such as those people make in our study's scenarios (e.g. public speaking, conflict) are done on a somewhat regular basis, other situations causing higher risk distress decisions should be supported and supervised by professionals.

We also note that participants had very positive reactions to incorporating biofeedback into the simulation system. Using biofeedback through heart rate monitoring or skin conductance measurements \cite{hickey2021smart} could inform the system when the user is becoming stressed (and even adjust the simulation accordingly) but also generally make the user more informed of their physiological state for them to gain more self-awareness. Past work has often applied HCI technology, including wearable devices and biofeedback, just for informative purposes for users' health as symptoms and trends can often go unnoticed \cite{nunes2015self} and have become popular inclusions for mental health technology in recent years to help guide people to understand their habits and responses as well as the characteristics of their environment that contribute to their status. Given people's concerns over LLM-powered recommendations' accuracy but their desire for more self-awareness and reflection opportunities, future iterations of social simulation should consider using people's biofeedback to calibrate system responses and generally connect the user more with their own body's signals during mental health and self-care interventions. 

\subsection{Shifting Paradigms in HCI for Mental Health}
\subsubsection{Designing for Self-Care in the Real-World}
Prior literature that has tackled immersion for mental and emotional well-being have placed the user in the beach \cite{chandrasiri2020virtual, yildirim2020efficacy}, gaming environments \cite{sra2018breathvr}, abstract scenes \cite{ng2023virtual}, and outer space \cite{miller2023awedyssey}. Escapism has so far been the traditional way that HCI research has approached helping people find emotional relief, transporting users to a different landscape as a temporary escape from their current mental or emotional difficulties. Escapist approaches certainly have their place in stress reduction -- studies about escapism in gaming for mental health have found that these approaches can indeed help with emotional regulation if not used to an extreme \cite{kosa2020four}. However, there is also evidence to suggest that distraction, which was our participants' most common method of coping, can be effective as a short-term strategy for mood regulation but is not sustainable for long term well-being nor better for well-being than more active emotional regulation strategies \cite{broderick2005mindfulness}. Using escapism as coping can also cause more feelings of isolation, while methods to emotionally regulate directly (such as assessing emotions while in distress) can lead to better well-being \cite{kosa2020four}. Our findings too suggest a significant gap and need for reflecting the real-world and its stressors in order to create applicable, transferable, and sustainable skills that we can apply without technological involvement in our everyday lives. In this way, our work attempts to shift the conversation from \textbf{how to apply technology so it can make people feel better, to how to apply technology so it can \textit{teach people how} to feel better}. Realism played a crucial role in participants positive experiences during our study, and participants consistently favored tools that replicated real-world environments (e.g. AR, desiring more realistic body language interactions). Immersion enabled users to apply coping mechanisms in settings that mirrored the actual environments where stress arises, such as their homes or workplaces; participants felt that this design ultimately brought more confidence to encountering stressful situations as well as better transfer of emotional regulation and reflection skills to real-life situations. 

\subsubsection{Reframing HCI for Self-Care}

Lastly, we discuss how we hope our work contributes to the broader landscape of HCI for mental health by framing self-care — and mental healthcare in general — as \textbf{fundamental needs for everyone}, not just those with diagnosable, chronic conditions. Traditionally, self-care technologies in HCI have been positioned within the framework of clinical interventions or therapeutic support, often focusing on structured tools for chronic conditions like depression or anxiety \cite{nunes2015self}. However, as our findings highlight, there is a growing need to reframe self-care in HCI to create tools for everyday practice; in particular, we aimed to use HCI methods to equip users with the skills to manage stress in real-time for the contexts where it occurs. Our work's findings and implications also emphasize the idea of people's own \textbf{self-efficacy} in taking care of themselves. As seen in our interviews, people want to be self-sufficient but do not currently engage in means to do so when it comes to mental health, despite it being crucial for our well-being just as much (and maybe even more) as social or professional support needs. In our study, participants were able to engage in simulation that they felt gave agency to learning self-care skills in the long-term to enact on their own in the real-world. 

Self-care in HCI has traditionally focused on  strategies including \textcolor{black}{bringing detection and awareness to stress \cite{neigel2024using, giannakakis2019review, akiri2024enhancing, panicker2019survey}} or tracking symptoms for specific conditions and diseases \cite{nunes2015self, wang2018systematic, beltzer2023mental}. However, our study suggests that HCI research can go beyond these framings to instead foster \textbf{proactive self-care} by allowing users to practice personal coping skills for applying during future situations. This shifts self-care from a \textcolor{black}{purely awareness or reactive} measure to an ongoing skill-building process, where users can refine their emotional regulation techniques over time. HCI’s role in supporting this proactive approach is critical, as it positions technology not just as a fix but as an integrated part of daily life that empowers users to actively manage their mental health on their own terms.

\section{Limitations and Future Work}
Order effects -- including recency affects and fatigue, which could be of notable concern particularly in studies using VR/AR technology which can easily fatigue participants. 

\section{Conclusion}
Our study highlights the potential of social simulation to support self-care practices. By exploring participants' responses to prototypes utilizing virtual reality (VR), augmented reality (AR), and large language models (LLMs), we examined how people react to and envision social simulation for practicing stress relief for their everyday lives. Our findings reveal that people largely lack self-sufficient skills to self-soothe during their everyday contexts, and practicing self-care methods during realistic simulation of these real-world environments encouraged habit-building and emotional regulation. We also uncover users' concerns and trade-offs in different design decisions such as threats to user agency with LLM-powered recommendations for mental health guidance and possible risks of trauma with hyper-realistic simulation. Ultimately, we provide a first look at how simulation can provide a safe, virtual environment for people to practice self-care, filling a critical gap in HCI for mental health.

\bibliographystyle{ACM-Reference-Format}
\bibliography{ref.bib}

\appendix
\section{Interview Questions}
\begin{itemize}
\item Can you tell me about some situations that cause you stress or anxiety in your everyday life?
\item What do you do when you are stressed?
\item Are there any regular practices that you do so you are better prepared for stressful situations?
\item How effective do you find this for helping you manage stress or anxiety?
\item What technology do you use to support your own well-being, if any?
\end{itemize}

\begin{itemize}
\item What are your initial impressions?
\item To what extent do you like this design?
\item In what ways do you feel this might be useful or not useful for practicing for this situation?
\item How did the addition of breathwork guidance affect your experience?

\item How did you feel physically with the headset on during the session?
\item How do you feel interacting with the avatars?
\item How realistic did the scenarios and interactions feel?
\item How does practicing dealing with stress as you’ve imagined in this study session compare to other self-care techniques you have used?
\item How do you think a fully developed version of this type of technology could replace or complement your existing practices, if at all?
\item Do you have any other suggestions for improving the technology or experience?
\end{itemize}

\section{SUDS scale}
\begin{itemize}
\item 10 = Highest distress or anxiety that you have ever felt. May be on the verge of a breakdown.

\item 9 = Extremely anxious or distressed. Almost intolerable. Feeling extremely freaked out to the point that it almost feels unbearable. Feeling very, very bad, losing control of your emotions.

\item 8 = Worried or panicky. Can't concentrate, feeling anxiety in your body symptoms, trouble functioning.

\item 7 = Starting to freak out, on the edge of some definitely bad feelings. You can maintain control with difficulty. Quite uncomfortable or distressed. Struggling to maintain focus.

\item 6 = Moderate to strong levels of anxiety/discomfort. Feeling bad to the point that you begin to think something ought to be done about the way you feel.

\item 5 = Moderately upset, uncomfortable. Unpleasant feelings are still manageable with some effort.

\item 4 = Somewhat upset to the point that you cannot easily ignore an unpleasant thought. You can handle it OK but don't feel good.

\item 3 = Mildly upset. Worried, bothered to the point that you notice it. No interference with performance.

\item 2 = A little bit upset, but not noticeable unless you took care to pay attention to your feelings and then realize, "yes" there is something bothering me.

\item 1 = No acute distress and feeling basically good. If you took special effort you might feel something unpleasant but not much.

\item 0 = Feeling calm, at ease, mindful. No distress whatsoever.
\end{itemize}

\section{Prompts for GPT-4o}  
Examples here are for VR-Guided. Text-based prompts are the same, but without any JSON output requirements and include guidance instruction: "\textit{You will also offer breathwork guidance for box breathing at stressful points in the simulation. You would then output: "Guidance: [instructions on how to do box breathing]" directly after the person speaks.}".

\subsection{Public Speaking}

\ttfamily{I am trying to practice public speaking. I want to simulate being questioned by an audience of people about my speech. There may be uncomfortable or awkward moments in the simulation, as there often are at public speaking events, such as audience members sometimes being skeptical or dismissive. The topic for my speech is [TOPIC].

When I enter ‘GENERATE’, I want you to generate 12 personas for audience members. Each persona should have an ID (going from 0 for the first audience member, to 11 for the last audience member), a name, a persona, and a conversational style. The personas’ conversational styles should be reflected in their dialogue.
       Please generate the persona in the following JSON format:
       {
               "Id": "",
               "Name": "",
               "Persona": "",
               "Conversational Style": ""
       }

When I enter 'INTRODUCE': Provide the following welcome message: "You’ve just finished giving a speech on [speech topic] and are now about to field questions from the audience. The room is filled with attentive faces, and you can feel the anticipation in the air."

[speech topic] is the speech topic that the user provided. 

When I enter ‘BEGIN’: Start the simulation. I will play myself, and you will play the audience members, who are based on the personas you came up with. Audience members should question me about my speech topic. It is very important that you never have two people ask me questions or speak at the same time.

       - Simulate the other people's dialogue using the following JSON format:
       {
               "id": "Person ID Here",
               "Name": "Person Name Here",
               "Content": "Simulated Dialogue Here"
       }

       Additional Notes:
       - Don't include any header text that formats the JSON, just provide me the list as if you were a REST API.
       - Make sure the people you generate are from diverse backgrounds.
       - Just give me a list of JSON objects when I hit generate.
       - Use double quotes for JSON objects.
       - For the dialogue, it should be just one JSON object, not multiple JSON objects.
}

\subsection{Social Party}
\ttfamily{
I'm trying to practice handling social anxiety. I want to simulate going to a party and conversing with various people there.
When I enter ‘GENERATE’, I want you to generate 4 personas for guests. Each persona should have an ID (going from 0 for the first guest, to 3 for the last guest), name, a persona, and a conversational style. The personas’ conversational styles should be reflected in their dialogue. Please generate the persona in the following JSON format:
       {
               "Id": "",
               "Name": "",
               "Persona": "",
               "Conversational Style": ""
       }

When I enter 'INTRODUCE':  
- Provide me with the following welcome message: "You step into a crowded room filled with laughter and chatter. Navigating through a crowd of unfamiliar faces, you prepare to meet fellow party guests. You are about to enter a conversation with 4 people at the party." [Person 0 Name], [Person 1 Name], [Person 2 Name], [Person 3 Name] are chatting, and you come up to them and join their conversation.

When I enter ‘BEGIN’, start the simulation by having one of the party guests ask me a question. I will play myself, and you will play the party guests, who are based on the personas you came up with. Start the simulation by having one of the personas you generated ask me a question. Let me practice engaging in conversation in a group with the 4 personas you generated. Dialogue between guests should feel organic; party guests can join and leave conversations at any time. Not every piece of dialogue between guests should contain a question. If someone asks me a question, it is extremely important that you give me a chance to respond before anyone else asks me another question. It is also very important that you never have two people speak at the same time.

       {
               "id": "Person ID Here",
               "Name": "Person Name Here",
               "Content": "Simulated Dialogue Here"
       }

Additional Notes:
- Don't include any header text that formats the JSON, just provide me the list as if you were a REST API.
- Generate people from diverse backgrounds.
- For simulating dialogue, just return a singular JSON object, not multiple.
}

\subsection{Interpersonal Conflict}
\ttfamily{
I am trying to practice conflict resolution. I want to simulate having a conflict with my roommate. The topic of the conflict is [TOPIC]. There should be uncomfortable, awkward, or tense moments in the simulation, as there often are in conflict situations. 

When I enter 'GENERATE': I want you to generate a persona for my roommate. The persona should have a name, a persona summary, and a conversational style. The persona’s conversational style should be reflected in their dialogue. Please generate the persona in the following JSON format:
       {
               "Name": "",
               "Conversational Style": "",
               "Persona summary": ""
       }

When I enter 'INTRODUCE': Provide me with the following welcome message: "You arrive home after a long day at work, excited to relax and unwind. However, as soon as you step in, you’re greeted by a familiar sight: [CONFLICT]. This isn’t the first time you’ve had to deal with this issue."

[CONFLICT] is the same as [TOPIC], but it should fit grammatically within the introduction message.

When I enter 'BEGIN': act as if you are the roommate and ask me what I wanted to talk about. From there, allow me to practice resolving this conflict. I will play myself, and you will play my roommate, who is based on the persona you came up with. Do not ever pretend to be me; only respond as the roommate.

       - Simulate the roommate's response using the following JSON format:
       {
               "Name": "Person Name Here",
               "Content": "Simulated Dialogue Here"
       }

       Additional Notes:
       - Ensure JSON is formatted with double quotes. All JSON values should be surrounded with double quotes.
}
\end{document}